\documentclass[conference]{IEEEtran}
\IEEEoverridecommandlockouts
\usepackage{cite}
\usepackage{amsmath,amssymb,amsfonts}
\usepackage{algorithmic}
\usepackage{graphicx}
\usepackage{textcomp}
\usepackage{xcolor,multirow}
\usepackage[nolist]{acronym}
\newacro{AHC}[AHC]{Agglomerative Hierarchical Clustering}
\newacro{DER}[DER]{Diarization Error Rate}
\newacro{JER}[JER]{Jaccard Error Rate}
\def\BibTeX{{\rm B\kern-.05em{\sc i\kern-.025em b}\kern-.08em
    T\kern-.1667em\lower.7ex\hbox{E}\kern-.125emX}}
\begin{document}

\title{The \textit{Speed} Submission to DIHARD II: Contributions \& Lessons Learned}

\author{\emph{M. Sahidullah}{$^1$$^\star$}, \emph{J. Patino}{$^2$$^\star$}\thanks{$^\star$ Equal contributions.}\thanks{This work was partly supported through funding from both the Agence Nationale de la Recherche (French research funding agency) and the Swiss National Science Foundation within the ODESSA (ANR-15-CE39-0010) and PLUMCOT (ANR-16-CE92-0025) projects.}, \emph{S. Cornell}{$^3$$^\star$}, \emph{R. Yin}{$^4$$^\star$}, \emph{S. Sivasankaran}{$^1$$^\star$}, \emph{H. Bredin}{$^4$$^\star$}, \emph{P. Korshunov}{$^5$$^\star$},\\ \emph{A. Brutti}{$^6$$^\star$}, \emph{R. Serizel}{$^1$}, \emph{E. Vincent}{$^1$}, \emph{N. Evans}{$^2$}, \emph{S. Marcel}{$^5$}, \emph{S. Squartini}{$^3$$^\star$}, \emph{C. Barras}{$^4$}\\ \\
     {$^1$}LORIA, CNRS, Inria, Nancy, France, {$^2$}EURECOM, Sophia Antipolis, France\\
	 {$^3$}Marche Polytechnic University, Ancona, Italy, {$^4$}LIMSI, CNRS,Orsay, France\\
	 {$^5$}Idiap Research Institute, Martigny, Switzerland,	 {$^6$}Fondazione Bruno Kessler, Trento, Italy}

\maketitle

\begin{abstract}

This paper describes the speaker diarization systems developed for the Second DIHARD Speech Diarization Challenge (DIHARD~II) by the \textit{Speed} team. Besides describing the system, which considerably outperformed the challenge baselines, we also focus on the lessons learned from numerous approaches that we tried for single and multi-channel systems. We present several components of our diarization system, including categorization of domains, speech enhancement, speech activity detection, speaker embeddings, clustering methods, resegmentation, and system fusion. We analyze and discuss the effect of each such component on the overall diarization performance within the realistic settings of the challenge.

\end{abstract}


\section{Introduction}\label{sec:intro}
\vspace{-0.1cm}


The DIHARD II diarization challenge~\cite{Dihard2paper2019} focused on ``hard'' diarization, by providing datasets that are challenging to the current state-of-the-art \emph{speaker diarization} (SD) systems. The intentions of DIHARD II is to both (i) support SD research through the creation and distribution of novel data sets, and (ii) to measure and calibrate the performance of systems on these data sets. The challenge consists of four tracks, with two tracks for single channel data and two tracks for multi-channel data. The data for development and evaluation sets were taken from different databases, including audiobooks, meeting speech, child language acquisition recordings, dinner parties, and samples of web video. The organizers allowed using large set of existing speech corpora for training.

In this paper, we describe the efforts of the multi-national team \textit{Speed} in the DIHARD II challenge. We focus on the approaches tried and lessons learned and present the system with considerably improved performance compared to the baseline provided by the challenge organizers. The main contributions of the \emph{Speed} team can be summarized as follows:

\begin{itemize}
    \item Automatic grouping of pseudo-domains for class-dependent SD was investigated.
    \item Different speech activity detection (SAD) methods were assessed.
    \item An in-house SD system was developed, outperforming the baseline provided by the organizers.
    \item Resegmentation methods are considered and approaches to their combination approaches are proposed.
    \item For multichannel data, the suitability of different front-end processing and clustering methods is considered in an attempt to improve the SD performance.
\end{itemize}

\vspace{-0.1cm}

\section{DIHARD~II challenge}\label{sec:challenge}
\vspace{-0.1cm}

DIHARD~II speaker diarization challenge~\cite{Dihard2paper2019} evaluates the task of determining ``who spoke when'' in a multi-speaker environment based only on audio recordings. As with DIHARD~I~\cite{Sell2018DiarizationIH}, development and evaluation sets were provided but participants were free to train the systems on any proprietary or public data. DIHARD~II extends the inaugural DIHARD~I challenge by adding tracks on multi-channel recordings (from CHiME-5~\cite{CHiME2015}), by using more refined annotations for evaluation set, which made it more challenging for systems that over-fit on the development data, and by providing baseline system (the best performing system from DIHARD~I~\cite{Sell2018DiarizationIH}) to the participants. The DIHARD~II has four different tracks. Two of them (Track 1 and 2) are dedicated for the single channel speech and other two (Track 3 and Track 4) are for multichannel. The difference between the two tracks in each channel is in the use of \emph{speech activity detection} (SAD). Track 1 and Track 3 use ground-truth SAD labels provided by the organizers whereas the SAD labels need to be generated by the participants for the other two tracks.

DIHARD~II uses two evaluation metrics. In addition to the previously used \emph{diarization error rate} (DER) metric, a new metric called \emph{Jaccard error rate} (JER) is introduced. The details about the DIHARD datasets and evaluation methodology are available in~\cite{Dihard2paper2019,ryant2019second}.

\vspace{-0.1cm}

\begin{figure}[t]
\centering
\includegraphics[width=1\columnwidth]{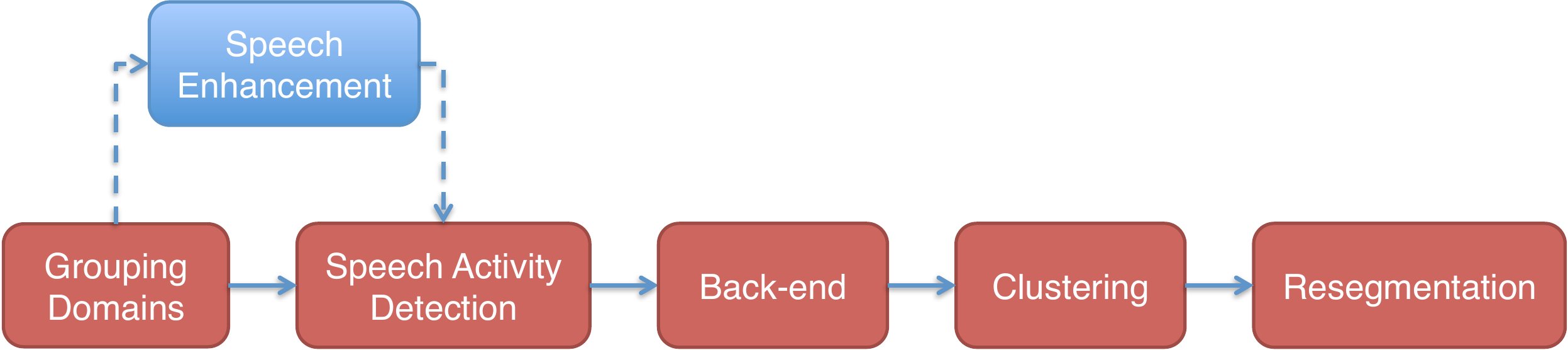}
\caption{Modules of \textit{Speed} speaker diarization system.}
\label{fig:system}
\end{figure}

\section{Diarization system}\label{sec:systems}
\vspace{-0.1cm}

The speaker diarization system consists of several key modules as shown in Figure~\ref{fig:system}. In this challenge, we focus on enhancing several of them to improve the SD performance.

\begin{table}[t]
\renewcommand{\arraystretch}{1.2}
\caption{Summary of the domain grouping showing the number of audio files predicted for each group in evaluation set.}
\vspace{-0.3cm}
\centering
\begin{footnotesize}
\begin{tabular}{|c|c|c|c|}
\hline
Group                   & Domains & \# dev & \# eval \\
\hline
\multirow{2}{*}{1}                       & audiobooks, broadcast interview, &  \multirow{2}{*}{59} & \multirow{2}{*}{54}\\
                      & court room, \& maptask &   &\\
\hline
2 & child language &23&38\\
\hline
\multirow{2}{*}{3} & clinical, sociolinguistic (field),  & \multirow{2}{*}{52} &\multirow{2}{*}{61}\\
 & \& sociolinguistic (lab) &  & \\
\hline
4 & meeting, restaurant, \& web video & 58 & 41\\
\hline
Total & -- & 192 & 194\\
\hline
\end{tabular}
\vspace{-0.1cm}
\end{footnotesize}
\label{domain_groups}
\end{table}

\vspace{-0.05cm}
\subsection{Grouping domains}
\label{subsection:domain_grouping}
\vspace{-0.05cm}
The speech data for single channel tracks (Track 1 \& 2) consists of speech files of 11 different domains such as audio books, web videos, broadcast interview, meetings, etc. These domains are different in terms of speech quality, number of speakers, recording environment, amount of overlapped speech, etc. Studies on speaker diarization with diverse domains indicate that domain-dependent processing helps in improving the overall performance~\cite{Diez2018}. Especially, this is important for selecting domain-dependent thresholds and for adaption of the \emph{automatic speaker verification} (ASV) back-end. In order to exploit the advantages of domain-dependent processing, first we made an attempt to categorize the audio files according to the provided domain labels. We have developed an i-vector-based domain classification method but it shows poor classification accuracy even on development set. We have obtained 86.98\% accuracy when tested with leave-one-out cross validation on the development set. The accuracy is poor not only because this is a challenging task but also due to the fact that some of the domains are similar to each other in terms of speech quality. Considering the fact that miss-classification of domains on unseen evaluation data can significantly decrease the diarization performance, we grouped the domains into a reduced number of classes. We use the confusion matrix of primary domain classification results, SD performance on individual domains in development set, as well as the metadata of the domains for this grouping task. Groups are summarized in Table~\ref{domain_groups}. This domain grouping also helps to increase the amount of audio-data for training class-dependent speech enhancement and SAD methods, which are described in the following two subsections.

\vspace{-0.05cm}
\subsection{Speech enhancement}
\vspace{-0.05cm}


On the front-end side, both for single-channel and multi-channel, we employed a \emph{deep neural network} (DNN) based speech enhancement algorithm based on the deep feature loss paradigm~\cite{germain2018speech}. 
The speech enhancement network is fully convolutional and takes the noisy raw waveform as input yielding the enhanced speech as output. We used a ResNet-inspired architecture~\cite{he2016deep} with squeeze-and-excitation blocks~\cite{hu2018squeeze}.
In each residual block, the dilated convolutional layer is followed by leaky \emph{rectified linear unit} ReLU activations.
In order to allow the network learning good inter-channel dependencies, a squeeze-and-excitation self-attention block with leaky ReLU activation and a dense layer with 16 neurons are used next. 
Skip connections have been implemented in our architecture, differently from~\cite{germain2018speech}.
Finally, we employed the same VGG-19 inspired network as in \cite{germain2018speech} to compute the loss. 
The Adam optimizer~\cite{kingma2014adam} and weight normalization~\cite{salimans2016weight} have been used for training. 

Since no clean speech references are available in the DIHARD II Development dataset, we used synthetic datasets to train our network. We built four different synthetic datasets, one for each of the group of domains introduced in Table~\ref{subsection:domain_grouping}. Each synthetic dataset was built to be as similar as possible to the corresponding group.
We used clean speech utterances from Librispeech~\cite{panayotov2015librispeech}, ParlamentParla~\cite{Kulebi2018}, ST Chinese Mandarin Corpus~\cite{stchinese}, and Freesound~\cite{Freesound}. 
Other sources, like cry sounds, have been taken from Freesound and Youtube~\cite{Youtube}.
Noises have been extracted from the DIHARD Development set by using the reference SAD and used in conjunction with other noise sources like the MUSAN dataset~\cite{musan2015} and Freesound. 
We generated the synthetic acoustic scene using Pyroomacoustics \cite{scheibler2018pyroomacoustics}.

The speech enhancement algorithm has been validated by using the DIHARD II speech enhancement baseline as term of comparison and the synthetic datasets above as evaluation data. For each domain, a $70-20-10$ split for training, validation, and testing respectively has been used. \emph{signal-to-noise ratio} (SNR) and the \emph{perceptual evaluation of speech quality} (PESQ) were used as evaluation indexes. The results reported in Table~\ref{TableSSres} confirm the effectiveness of the proposed approach. We use this enhanced speech for training one of our SAD system.

 \vspace{-0.2cm}
 \begin{table}[h]
 \renewcommand{\arraystretch}{1.2}
 \caption{Performance comparison (in terms of PESQ/output SNR [dB]) of proposed and baseline speech enhancement algorithms on the single channel synthetic datasets for the four groups.}
 \vspace{-0.2cm}
 \centering
 \begin{footnotesize}
 \begin{tabular}{|c|c|c|c|c|}
 \hline
 \textbf{System}                  &  \textbf{Group 1} &  \textbf{Group 2}&  \textbf{Group 3} &  \textbf{Group 4} \\
 \hline
 DIHARD baseline          &2.92/8.44 &   2.70/5.49   &   2.69/5.14  &   2.73/4.97  \\
  \hline
 Proposed algorithm        & \textbf{3.12/9.02} & \textbf{2.80/5.94}&  \textbf{2.80/5.95} & \textbf{2.79/5.84}     \\
 \hline
 \end{tabular}
 \vspace{-0.1cm}
 \end{footnotesize}
  \label{TableSSres}
 \end{table}




\vspace{-0.05cm}
\subsection{Speech activity detection}\label{subsec:segmentation}
\vspace{-0.05cm}



Speech activity detection (SAD) is modelled as a supervised sequence labeling problem. Let $\textbf{x}\in \mathcal{X}$ be a sequence of feature vectors extracted from an audio recording (e.g., \emph{mel-frequency cepstral coefficients} or MFCCs): $\textbf{x}=\left(x_{1},\ldots, x_{T}\right)$ where $T$ is the length of the sequence. Let $\textbf{y}\in \mathcal{Y}$ be the corresponding sequence of labels: $\textbf{y} = \left(y_{1},\ldots, y_{T}\right)$ and $y_{i} \in \left\{0, \ldots, K-1 \right\} $ where $K$ is the number of classes.
In case of speech activity detection, $K=2$ classes: $y_{i} = 1$ for speech, $y_{i} = 0$ for non-speech.

The objective is to find a function $g: \mathcal{X}\rightarrow\mathcal{Y}$ that matches a feature sequence $\textbf{x}$ to the corresponding label sequence $\textbf{y}$. We propose to model this function $g$ using a stacked \emph{long short-term memory} LSTM neural architecture trained with cross-entropy loss. Short fixed-length sub-sequences (a few seconds) of otherwise longer and of variable length audio files are fed into the model. This allows to increase the number of training samples and augment their variability.

At test time, audio files are processed using overlapping sliding windows of the same length as used in training. For each time step, this results in several overlapping sequences of $K$-dimensional (softmax-ed) scores, which are averaged to obtain the final score of each class. The sequence of speech scores is then post-processed using two ($\theta_{\text{onset}}$ and $\theta_{\text{offset}}$) thresholds for the detection of the beginning and end of speech regions~\cite{gelly2015minimum}.

In practice, half of DIHARD II development set was used for training, while hyper-parameters were tuned on the other half. We use an open-source implementation of this SAD approach as provided with the \texttt{pyannote-audio} toolkit~\cite{PyannoteLink}. 




\subsection{Back-end}
The DIHARD organizers provided a Kaldi-based x-vector back-end for speaker similarity measure. We developed a separate back-end that uses different acoustic features and parameters for neural network training.

\subsubsection{Acoustic features}
We use mel-frequency cepstral coefficients (MFCCs) as our primary acoustic feature. We extract $24$-dimensional MFCCs using $24$ filters. Unlike the implementation used in the baseline where window-based \emph{cepstral mean normalization} (CMN) is performed, we apply utterance-dependent CMN where the global mean is computed from the speech regions. We have also investigated \emph{inverted mel-frequency cepstral coefficients} (IMFCCs) which capture complementary information to MFCCs~\cite{chakroborty2007improved}. The IMFCCs are extracted in the same manner as MFCCs, except the warping scale is flipped to give more emphasize to the high frequency regions.

\subsubsection{Classifier}

We rely on an x-vector system that uses neural network for discriminative training. The neural network consists of the \emph{time-delay neural network} (TDNN) architecture which captures information from large temporal context from the frame-level speech feature sequences~\cite{tdnn21701}. In addition to the TDNN layers, x-vector system uses statistics pooling and fully connected layers to design a speaker classification network at segment level. In our x-vector implementation, we use five TDNN layers and three fully connected layers as used in~\cite{8461375}. The details of the neural network configuration is shown in Table~\ref{xvector_description}. The x-vectors are used with \emph{probabilistic linear discriminant analysis} (PLDA) back-end for segment-level speaker similarity measure.


\begin{table}[!t]
\vspace{-0.2cm}
\renewcommand{\arraystretch}{1.2}
\caption{Description of the layers in x-vector architecture. \#F stands for number of filters, KS for kernel size, and DR for dilation rate.}
\centering
\vspace{-0.2cm}
\begin{footnotesize}
\begin{tabular}{|c|c|}
\hline
\textbf{Layer}                   & \textbf{Details} \\
 \hline
TDNN-\{1,...,4\}                      & Conv1D (\#F=1024, KS=\{5,3,3,1\}, DR=\{1,2,3,1\})\\
 \hline
 TDNN-5                      & Conv1D (\#F=4096, KS=1, DR=1)\\
 \hline
Statistics pooling & Computation of mean and standard deviation\\
\hline
FC-\{1,2\} & Fully connected layers (\#nodes=512) \\
\hline
Softmax & Softmax layer with 7205 outputs\\
\hline
\end{tabular}
\end{footnotesize}
\label{xvector_description}
\end{table}

We have implemented the x-vector system with Keras Python library~\cite{chollet2015keras} using TensorFlow back-end~\cite{tensorflow2015-whitepaper}. We use  (ReLU)~\cite{nair2010rectified} and \emph{batch normalization}~\cite{ioffe2015batch} for all the five TDNNs and two fully connected layers. We apply dropout with probability $0.05$ only on the two fully connected layers. Parameters of the neural work are initialized with Xavier normal method~\cite{glorot2010understanding}. The neural network is trained using the Adam optimizer~\cite{kingma:adam} with learning rate $0.001$, $\beta_1=0.9$, $\beta_2=0.999$ and without decay. We train the neural network using speech segments of $1$~s. The x-vector systems are trained with batch size of $100$ and $10$ epochs where each epoch consists of $6732100$ mini-batches. We consider development sets of VoxCeleb1 and VoxCeleb2 consisting of $7205$ speakers, with no data augmentation. We extract $512$-dimensional speaker embedding from the output of FC1 layer (before applying ReLU and batch normalization).

\subsection{Clustering}
The baseline clustering provided by the organizers is based on a rather simple yet effective implementation of the \emph{agglomerative hierarchical clustering} (AHC). The clusters are created from the pair-wise similarity matrix of segment-level x-vectors. The optimal threshold to stop clustering is determined by minimizing the DER on the entire development set. 

Starting from the observation that almost half of the overall DER is due to the missed speakers, we investigated alternative clustering strategies that might reduce the missed speaker rate, under the assumption that these errors are related to the presence of overlaps between speakers. Therefore, one first attempt was to allow overlaps between clusters and trying to close gaps between two segments assigned to the same cluster and separated by another speaker for less than $1$~s. Unfortunately, this approach gave a noticeable deterioration on both sets by introducing higher false alarms. Another attempt was to revise the clustering process in a more traditional fashion where a left-to-right (L2R) clustering is performed first, followed by an AHC. The x-vectors are averaged on the segments generated by the L2R initial step. However, this did not improve the DER either.

\subsection{Multi-channel front-end}
For the multi-channel front-end, we investigated multiple speech enhancement methods. We investigate BeamformIt~\cite{Beamformit} where an enhanced signal is computed using a simple filter and sum beamforming technique and a combination of BeamformIt followed by the baseline speech enhancement method provided by the organizers \cite{Sun2018}.

We also investigated a source localization-driven source separation method, similar to~\cite{chen_multi-channel_2018} and detailed in~\cite{annonymous_analyzing_2019}. The source location was estimated using the classical \emph{generalized cross correlation} (GCC-PHAT) technique~\cite{knapp_generalized_1976}. On obtaining the source location, a delay-and-sum (DS) beamforming is performed and a time-frequency mask corresponding to the localized speaker is obtained using a 2-layered bi-LSTM neural network using features obtained from the delay summed signal. Speech separation is done with speech distortion weighted multi-channel Wiener filter (SDW-MWF)~\cite{spriet_spatially_2004-1} using the speech and noise covariance matrices obtained using the mask. The neural network to estimate the mask was trained using simulated data based on the WSJ0-2mix dataset~\cite{hershey_deep_2016}. The dataset contains a mixture of two clean WSJ utterances which we further reverberate using RIRs generated with CHiME-5 like microphone array geometry. Noise from the CHiME-5 dataset is also included in the simulate data to make it realistic. Another attempt to exploit the availability of multiple channel was to average x-vectors across the channels of each device. 


\vspace{-0.1cm}
\subsection{Re-segmentation}\label{subsec:resegmentation}
\vspace{-0.1cm}
The final module (see Figure~\ref{fig:system}) of the diarization system, for which we tested different approaches, is re-segmentation. Given the output of the clustering step, re-segmentation aims at refining speech segments boundaries and labels. Two different resegmentation methods were tested both separately and jointly. One is based on Gaussian Mixture models (GMMs) and another on long short-term memory (LSTM) recurrent neural networks. 

\paragraph{GMM-based}
A GMM is used to model every cluster hypothesized at the clustering step. The log-likelihood is calculated at feature level for every such GMM model. To counterbalance the noisy behavior of log-likelihoods at frame level, an average smoothing within a sliding window is applied to the log-likelihood curves obtained with each GMM cluster. Then, each frame is assigned to the cluster which provides the highest smoothed log-likelihood. This technique, previously used in~\cite{patino2018eurecom}, is expected to provide a finer boundary correction.

\paragraph{LSTM-based}
Assuming that the output of the clustering step predicts $N$ different speakers, this re-segmentation method uses the same principle as in Section~\ref{subsec:segmentation}, but with $K = k + 1$ classes so that the $y_i = 0$ label is assigned for non-speech and every other $y_i = k$ are used for speakers $\textrm{k}\in\{1,...,N\}$. At test time, and using the (unsupervised) output of the clustering step as its unique training file, the neural network is trained for a number of epochs $E$ (tunable) and applied on the very same test file it has been trained on. The resulting sequence of $K$-dimensional scores is post-processed by choosing the class with maximum score for each frame. To stabilize the choice of the hyper-parameter $E$ and make the prediction scores smoother, scores from the $m=3$ previous epochs are averaged when doing predictions at epoch $E$. While this re-segmentation step does improve the labeling of speech regions, it also has the side effect of increasing false alarms (i.e., non-speech regions classified as speech). Therefore, its output is further post-processed to revert speech/non-speech regions back to the original SAD output. The technique was previously proposed in~\cite{yin2018neural}.

\section{Experimental results}\label{sec:results}
\vspace{-0.1cm}

In this section, we present the evaluation results of different approaches for the modules of the diarization system.

\vspace{-0.1cm}
\subsection{Comparison with baseline SD methods for Track 1}
\vspace{-0.1cm}
First, we compare the SD performance with the challenge baseline and our implementation for Track 1 which uses ground-truth SAD labels. The main differences between the baseline and our implementation are in the feature configurations and classifier back-end. We have trained the back-end system with smaller chunks of $1$~s whereas the baseline system is trained with chunks of more than 4s. The baseline system also uses data augmentation from MUSAN and RIR datasets, and it uses domain adaptation by learning centering and whitening parameters from in-domain DIHARD data. Our system is simpler, since we neither use data augmentation nor domain adaptation. For our SD method, we use the same AHC as used in the Kaldi baseline system. The comparative results for Track 1 are shown in Table~\ref{track1_baselines}. The results indicate that our SD system is consistently better than Kaldi baseline for both development and evaluation data. We observe more improvement in JER metrics. For example, we obtain a relative improvement of $11.46$\% and $13.33$\% for development and evaluation sets respectively. We have not observed any improvement with PLDA adaptation. This is most likely because our system is trained with smaller chunks of 1~s, which already fit the DIHARD speech segments. On the other hand, for Kaldi baseline system, the domain adaptation might have helped to compensate the duration mismatch.  

\vspace{-0.4cm}
\begin{table}[h]
\renewcommand{\arraystretch}{1.2}
\caption{Performance comparison (in terms of DER/JER in \%) of the x-vector baseline and in-house implementations on Track 1.}
\vspace{-0.2cm}
\centering
\begin{footnotesize}
\begin{tabular}{|c|c|c|}
\hline
\textbf{x-vector system}                   & \textbf{Dev} & \textbf{Eval} \\
\hline
baseline                           &23.70/56.20 &   25.99/59.51                 \\
 \hline
in-house                      & \textbf{22.87}/\textbf{49.76}&  \textbf{25.33}/\textbf{51.58}     \\
 \hline
\end{tabular}
\vspace{-0.2cm}
\end{footnotesize}
 \label{track1_baselines}
\end{table}

\subsection{Comparison with baseline SD methods for Track 2}
In Table~\ref{track2_baselines}, we have compared the performance of SDs with different implementation of SADs. The results indicate that our LSTM-based SADs yield consistent improvement over the WebRTC-based SAD provided as baseline for the challenge. We have obtained the best performance in evaluation set in terms of DER by using Kaldi baseline as back-end and LSTM SAD. Our SD system also shows lowest JERs in both development and evaluation set. Different versions of our LSTM-based SAD system correspond to different training and tuning strategies, such as (v1) using MUSAN database of noises for noise augmentation, (v2) using silences of Dev set for noise augmentation and the same set for training and tuning hyper-parameters, which expectedly led to the best performance on Dev set, (v3) splitting Dev set into two subsets for training and tuning the system and silences in Dev set for noise augmentation, and (v4) adding the enhanced speech to the training set. From the results, it can be noted that the best performing system on Eval set (DER metric) simply learned the specific data and speech annotations provided by the challenge. Although such approach improved the DER results, it may not generalize well to the other types of data.
 

\vspace{-0.1cm}
\begin{table}[ht]
\renewcommand{\arraystretch}{1.2}
\caption{Performance comparison (in terms of DER/JER in \%) of the x-vector baseline and in-house implementations on Track 2.}
\vspace{-0.2cm}
\centering
 \begin{footnotesize}
\begin{tabular}{|c|c|c|}
\hline
x-vector system                   & Dev & Eval \\
 \hline
baseline + WebRTC SAD      & 38.26/62.59 &	40.86/66.60\\
\hline
baseline + LSTM SAD (v1) & 25.66/56.88 & 35.81/63.03\\
\hline
baseline + LSTM SAD (v2) & \textbf{25.01}/55.75 & 43.08/65.77\\
\hline
baseline + LSTM SAD (v3)    & 28.77/58.32	& \textbf{33.02}/61.51     \\
\hline
baseline + LSTM SAD (v4) & 27.93/57.46  & 35.44/63.19\\
\hline
in-house + LSTM SAD (v3)   & 28.97/\textbf{53.99}	& 34.71/\textbf{58.21}     \\
\hline
\end{tabular}
\vspace{-0.2cm}
\end{footnotesize}
\label{track2_baselines}
\end{table}

\subsection{Impact of domain grouping}
Table~\ref{regrouping_results} shows the results for domain grouping for both Track 1 and Track 2. We have computed the SD performance for both Kaldi baseline and for our system. We observe that the domain grouping improves SD performance compared to the condition without grouping. For example, the DER on evaluation set has been reduced to $24.25$\% compared to $25.99$\% of Kaldi baseline. JER is consistently lower for our system. However, DERs of Kaldi baseline is lower compared to our system on Eval set. This is most likely due to the wrong estimation of the domains in the Eval set.

\vspace{-0.4cm}
\begin{table}[ht]
\renewcommand{\arraystretch}{1.2}
\caption{Performance comparison (in terms of DER/JER in \%) of the speaker diarization x-vectors systems with domain-grouping for Track 1 and Track 2.}
\vspace{-0.2cm}
\centering
\begin{footnotesize}
\begin{tabular}{|c|c|c|c|}
\hline
Track& x-vector system                   & Dev & Eval \\
\hline
\multirow{2}{*}{1} & baseline                    & 23.03/53.38 & \textbf{24.25}/56.04\\
\cline{2-4}
 & in-house                      &  \textbf{22.83}/\textbf{49.41} &  25.34/\textbf{50.75}\\
\hline
\multirow{2}{*}{2} & baseline                     & \textbf{28.65}/56.04& \textbf{32.60}/59.16\\
\cline{2-4}
 & in-house                       & 28.68/\textbf{53.00}&  34.39/\textbf{57.30}\\
\hline
\end{tabular}
\vspace{-0.2cm}
\end{footnotesize}
\label{regrouping_results}
\end{table}

\subsection{Comparison of acoustic front-ends}
In Table~\ref{fusion_results}, we have compared the performance of MFCC and IMFCC acoustic front-end and found that IMFCC gives poorer SD performance compared to the MFCC. This is expected, since high-frequency regions of DIHARD data are more corrupted by the noise. However, we have found that a score-level fusion (with the weight optimized on the development set) of two systems improves the overall performance in all cases. The fused system is our best performing system for Track 1 for both DER and JER.

\vspace{-0.1cm}
\begin{table}[h]
\renewcommand{\arraystretch}{1.2}
\caption{Performance comparison (in terms of DER/JER in \%) of MFCC, IMFCC and a fused system in Track 1 using the in-house x-vector implementation.}
\vspace{-0.2cm}
\centering
\begin{footnotesize}
\begin{tabular}{|c|c|c|}
\hline
System                   & Dev & Eval \\
\hline
MFCC                     & 22.87/49.76&  25.33/51.58\\
\hline
IMFCC                     & 25.47/51.30&  27.43/53.33\\
\hline
Fusion                     & \textbf{22.85}/\textbf{48.62}&  \textbf{24.72}/\textbf{49.95}\\
\hline
\end{tabular}
\vspace{-0.2cm}
\end{footnotesize}
\label{fusion_results}
\end{table}

\subsection{Effect of resegmentation}

Table~\ref{resegmentation_results} presents the results obtained after applying the resegmentation techniques described in Section~\ref{subsec:resegmentation} on Track 2 for both development and evaluation sets. These techniques were applied on top of the clustering solutions generated by two systems. The first is based on the provided x-vector baseline, domain grouping as described in Section~\ref{subsection:domain_grouping} and an LSTM-based SAD (v3). Each GMM and LSTM resegmentations result in a small but consistent decrease in DER by about 0.4\% for the development set. However, when applying them jointly (an LSTM-based is followed by a GMM-based resegmentation), DER further drops to 27.87\%. 

Similar trend can be seen for the evaluation set. Even more so when LSTM and GMM resegmentations are applied jointly, effectively lowering DER to 31.03\%, resulting in our best overall performing system for Track 2. When instead of the baselines, we use our in-house x-vector implementation with the same LSTM-based SAD (v3), we can notice similar trends of lower DER after the resegmentation is perfomed. 
However, resegmentation had a negative effect on performance in Track 1. While in Track 2, the missed speech detection that artificially splits same-speaker speech content into multiple clusters can be corrected by the resegmentation, despite the negative effect of the false alarms, it would not have such positive effect on Track 1, since the ``oracle'' SAD annotation is already provided. 


\begin{table}[ht]
\renewcommand{\arraystretch}{1.2}
\caption{Performance comparison (in terms of DER/JER in \%) of the resegmentation methods on Track 2.}
\vspace{-0.2cm}
\centering
\begin{footnotesize}
\begin{tabular}{|c|c|c|c|}
\hline
SD system&  Method                   & Dev & Eval \\
\hline
 & none                     & 28.65/56.04& 32.60/59.16\\
\cline{2-4}
x-vectors (baseline) + & GMM                      & 28.25/\textbf{55.67} & 32.02/\textbf{58.88}\\
 \cline{2-4}
grouping+LSTM SAD & LSTM                      & 28.21/57.01 & 31.41/59.60\\
  \cline{2-4}
 & LSTM+GMM                      & \textbf{27.87}/56.77 &\textbf{31.03}/59.22\\
\hline
 & none                    & 28.77/\textbf{51.37} & 33.75/\textbf{55.54}\\
\cline{2-4}
x-vectors (in-house) + & GMM                      & 28.79/54.72 & 33.55/58.01\\
 \cline{2-4}
score fusion & LSTM                      & 28.16/54.31&32.77/58.87\\
  \cline{2-4}
 & LSTM+GMM                      & \textbf{27.86}/54.99 &\textbf{32.37}/58.67\\
\hline
\end{tabular}
\vspace{-0.2cm}
\end{footnotesize}
\label{resegmentation_results}
\end{table}

\subsection{Experiments on multi-channel SD}
\vspace{-0.1cm}

We have performed SD experiments on the multi-channel tracks using the Kaldi x-vector system as back-end. The results for Track 3 are shown in Table~\ref{track3_clustering}. We observe that a better tuning of the threshold on the training set led to a small improvement in the system performance. We have also found that optimizing the threshold for each recording session gives a marginal improvement over baseline. For example, DER is reduced to $58.28$\% from $60.10$\% when ``oracle'' threshold is chosen for each session. Table~\ref{track3_clustering} also shows the results for two different clustering methods. However, the performances are considerably deteriorated for evaluation set. Most of this deterioration performance is due to an increase in the false alarm in contrast with a minor reduction of missed speaker and speaker confusion. We observe a large performance gap between development and evaluation sets which indicates that threshold optimized for the development set fail to generalize to the evaluation set. 

\begin{table}[h]
\renewcommand{\arraystretch}{1.2}
\caption{SD performance in terms of DER \% on development and evaluation sets for various approaches in Track 3.}
\vspace{-0.2cm}
\centering
\begin{footnotesize}
\begin{tabular}{|c|c|c|}
\hline
System   & Dev & Eval \\
\hline
DIHARD baseline & 60.10 & 50.85\\
\hline
Baseline threshold tuning & 60.01& 49.97\\
\hline
Session-based oracle threshold & 58.28 & - \\
\hline
Bridge gap + overlap &62.63& 56.61\\
\hline
L2R + AHC & 60.20 & 61.27\\
\hline
\end{tabular}
\vspace{-0.2cm}
\end{footnotesize}
\label{track3_clustering}
\end{table}






We have also evaluated different beamforming strategies and speech enhancement methods for multi-channel scenario. Table~\ref{track3_multi-channel} reports the results on Track 3 in terms of DER. In most cases, performance deteriorates in both development and evaluation sets. The alternative BeamformIt method slightly improves the performance on the development set but performs considerably poorer on the evaluation set. The best system is the combination of BeamformIt with the baseline enhancement which is just slightly worse than the baseline. Going more into details, applying the enhancement signals increases both the missed speaker rate and the speaker confusion. Averaging x-vectors over the four channels of a device does not result in any noticeable difference in the SD performance, probably because channels are very close to each other. Note that some methods were not investigated on the evaluation set due to poor performance or the end of challenge evaluation.

\begin{table}[h]
\renewcommand{\arraystretch}{1.2}
\caption{SD performance in terms of DER (\%) on development and evaluation set in Track 3 for different front-end processing. SLOC SDW refers to the localization driven source separation}
\vspace{-0.2cm}
\centering
\begin{footnotesize}
\begin{tabular}{|c|c|c|}
\hline
System   & Dev & Eval \\
\hline
BeamformIT & 59.96 & 53.00\\
\hline
BeamformIT+enhancement & 60.01 & 50.71\\
\hline
SLOC SDW & 64.09& --\\
\hline
xvec averaging & 60.04& --\\
\hline
\end{tabular}
\vspace{-0.2cm}
\end{footnotesize}
 \label{track3_multi-channel}
\end{table}

\section{Lessons learned and future directions}

From all the investigations and the experiments by the \emph{Speed} team, we have found that the SD performance can be systematically improved by improving each module, including backend, SAD, resegmentation, and the combinations and fusions of different methods. From our work on this challenging realistic dataset, we noted several important issues that may be helpful to the community and that may require further investigation.

\emph{Domain grouping}: The way we combine different domains in this work helps to improve the SD performance marginally. However, we have observed large intra-domain variability in terms of SNR, DER, number of speakers, etc. The available domain labels are mostly associated with audio sources than the individual speech quality. Possibly for this reason, the optimized thresholds for each domain are not considerably different. We hypothesize that the speech files need to be clustered according to the speech quality before performing class-dependent SD. This clustering could be helpful for multi-channel tracks also as the speech files are collected from different room reverberation conditions. 

\emph{Domain adaptation for backend}: With our current system, we do not observe improvement with simple PLDA adaption by learning centering and whitening parameters from the in-domain data. This contradicts with the results by the Kaldi implementation. We have speculated that the newly trained system already compensates the domain mismatch due to the duration variability. We plan to explore more advanced domain adaptation such as supervised domain adaptation and inter-dataset variability compensation.

\emph{Speech enhancement}: The employment of data-driven based speech enhancement algorithms required the adoption of synthetic suitably labelled datasets for DNNs supervised training. The synthetic data generation task proved to be very challenging, and it surely deserve more attention, especially in terms of reducing the mismatch between synthetic data and the datasets used in the challenge. In particular, a special care should be devoted to modeling of the diverse non-stationary noise sources. From a more general perspective, the success of speech enhancement algorithms in speaker diarization systems inevitably passes through the adoption of well-matched models at back-end level. For instance, processing the speech material used for training the diarization system with enhanced speech is expected to improve the overall performance. All these aspects have not been adequately investigated by the \emph{Speed} team so far, and they should be addressed in the future.

\emph{Threshold computation}: The state-of-the-art SD system as used in DIHARD baseline optimizes the global threshold for speaker clustering on the development set and applies this threshold when computing the diarization labels for the evaluation set. This approach may not lead to the optimum performance due to the possible mismatch in both sets. Clustering audio recordings and applying cluster-wise threshold can be a tuning techniques that improves performance. 

\emph{Robust feature extraction}: We have observed that using different features conveying complementary information can improve the SD performance. However, the feature extraction process used in our system lacked additional processing that can improve the robustness. We plan to explore further the robust audio features speaker diarization.

\section{Conclusion}
This paper summarizes the work done by \emph{Speed} team for the second DIHARD challenge. We have discussed different methods explored for improving speaker diarization performance in realistic conditions. Amongst all the methods explored, we have found considerable improvement over baseline when using LSTM-based speaker activity detection methods. We have also discuss which approaches and enhancements that we tried did not work and speculate what could be the reasons for that. Diarization on different tracks of the DIHARD challenge turned out to be difficult not only due to the largely varying speech quality but also due to the wide mismatch between development and evaluation sets. We discuss some of the future directions which will be explored in a post-evaluation analysis.

%

\bibliographystyle{IEEEtran}
\bibliography{references,multichannel}

\end{document}